\documentclass[pdflatex,sn-mathphys-num,iicol]{sn-jnl}% Math and Physical Sciences Numbered Reference Style

\usepackage{graphicx}%
\usepackage{multirow}%
\usepackage{amsmath,amssymb,amsfonts}%
\usepackage{amsthm}%
\usepackage{mathrsfs}%
\usepackage[title]{appendix}%
\usepackage{xcolor}%
\usepackage{textcomp}%
\usepackage{manyfoot}%
\usepackage{booktabs}%
\usepackage{algorithm}%
\usepackage{algorithmicx}%
\usepackage{algpseudocode}%
\usepackage{listings}%
\usepackage{siunitx}%

\begin{document}

\title[Denoising Speckle]{Enhancing Speckle Metrology with Diffusion Denoising in Photon-Starved Regimes}

\author[1,2,3]{\fnm{Admir} \sur{Bajraktarevic}}%\email{iauthor@gmail.com}
\equalcont{These authors contributed equally to this work.}

\author[1,2]{\fnm{Alexander C.} \sur{Trowbridge}}%\email{iiauthor@gmail.com}
\equalcont{These authors contributed equally to this work.}

\author[4,5]{\fnm{Saba N.} \sur{Khan}}%\email{iiiauthor@gmail.com}
\author[6]{\fnm{Anton} \sur{van den Hengel}}%\email{iiiauthor@gmail.com}
\author[4]{\fnm{Graham D.} \sur{Bruce}}%\email{iiiauthor@gmail.com}
\author*[1,2,3,4]{\fnm{Kishan} \sur{Dholakia}}\email{kishan.dholakia@adelaide.edu.au}

\affil*[1]{\orgdiv{Centre of Light for Life}, \orgname{School of Biological Sciences}, \orgaddress{\street{Adelaide University}, \city{Adelaide}, \postcode{5005}, \state{SA}, \country{Australia}}}

\affil[2]{\orgname{Institute for Photonics and Advanced Sensing}, \orgaddress{\street{Adelaide University}, \city{Adelaide}, \postcode{5005}, \state{SA}, \country{Australia}}}

\affil[3]{\orgname{Australian Research Council Industrial Transformation Training Centre for Current and Emergent Quantum Technologies}, \orgaddress{\street{Adelaide University}, \city{Adelaide}, \postcode{5005}, \state{SA}, \country{Australia}}}

\affil[4]{\orgdiv{SUPA}, \orgname{School of Physics \& Astronomy}, \orgaddress{\street{University of St Andrews}, \city{North Haugh, St Andrews}, \postcode{KY16 9SS}, \country{UK}}}

\affil[5]{\orgname{School of Engineering and Physical Sciences},\orgaddress{\street{
Heriot-Watt University}, \city{Edinburgh}, \postcode{EH14 4AS},\country{ Scotland, UK}}}

\affil[6]{\orgname{Australian Institute for Machine Learning}, \orgaddress{\street{Adelaide University}, \city{Adelaide}, \postcode{5005}, \state{SA}, \country{Australia}}}

% use {asbstract*} to suppress the copyright line. Copyright information will be added in production

\abstract{Laser speckle is a powerful tool for precision metrology that enables highly sensitive measurements of light sources and subtle environmental perturbations. Many applications require operation in photon-limited regimes, for example when using low-power illumination or in spectral regions where sensitive detectors are unavailable. In these conditions, the structured speckle pattern that encodes the signal becomes challenging to disentangle from measurement noise, severely degrading performance.  Here, we introduce a denoising framework to separate measurement noise from the underlying speckle structure in low-signal data. Using a hybrid pre-training and experimental fine-tuning strategy, the model is adapted using a small experimental dataset and integrates directly with existing speckle metrology pipelines. Applied to femtometre-scale wavelength sensing using an integrating sphere, the approach reduces root-mean-square error in low-signal conditions by up to 72\% and enables accurate reconstruction where conventional speckle metrology fails.}
%\textbf{This enables a reduced optical power or exposure time with only a limited experimental calibration dataset for fine-tuning when adapting to a new optical configuration}.\textcolor{red}{odd last sentence}


\maketitle
\clearpage

%%%%%%%%%%%%%%%%%%%%%%%%%%  body  %%%%%%%%%%%%%%%%%%%%%%%%%%
\section{Introduction}
Laser speckle arises from the interference of coherent light scattered by a rough surface or transmitted through a disordered medium. Although often treated as an unwanted artifact, speckle patterns encode phase-sensitive information that enables precise measurements of parameters such as wavelength, displacement, the azimuthal index of beams carrying orbital angular momentum and other physical quantities \cite{Metzger2017,britto2026ultra,facchin_measuring_2023,Facchin2024,aman2026speckle, trowbridge2025radiation,Silva2022SignalCarryingSpeckle}. In particular, highly scattering elements, such as an integrating sphere, can enhance sensitivity by increasing the optical path-length diversity and the number of independent interference contributions, producing speckle patterns that decorrelate strongly under small perturbations \cite{Metzger2017,britto2026ultra,FacchinRI2022,facchin_measuring_2023,Facchin_2021}. Integrating spheres have been shown to offer exceptionally high sensitivity for speckle metrology \cite{Facchin2024}. More broadly, speckle metrology has found applications in biomedical measurements, including optical coherence tomography and real-time monitoring of blood flow and coagulation \cite{hajjarian_technological_2021,heeman_clinical_2019,vaz_laser_2016,gong_optical_2016,tripathi_assessing_2014,schmitt_speckle_1999}.

%However, despite the highly reflective nature of the internal coating, the extensive multiple scattering of light within an integrating sphere significantly reduces the detected optical power. Compared with other commonly used scattering media such as multimode fibres or ground glass, integrating spheres therefore require a greater input power. To compensate for this loss, a higher incident laser power or a longer camera exposure time are typically required to maintain sufficient signal-to-noise ratio for speckle-based measurements. In practice, such an approach can constrain operation in photon-limited regimes and introduce additional challenges. This includes thermal loading, temporal drift of the speckle pattern, and reduced measurement fidelity at short exposure times \cite{Facchin_2021}. Consequently, these ultimately limit the accessible operating regime of speckle-based metrology, particularly in applications where illumination levels or exposure durations need to be minimised.

Speckle-based metrology is inherently subject to trade-offs between sensitivity, stability, and optical throughput, as increasing optical path-length diversity enhances measurement sensitivity but typically reduces signal levels and can introduce additional instabilities \cite{kumar2025performancelimitonchipspeckle, Facchin2024}. This interplay is a general characteristic of disorder-based speckle systems, as demonstrated across both bulk and integrated implementations. It is particularly pronounced in strongly scattering systems such as integrating spheres, which achieve high sensitivity through extensive multiple scattering but at the cost of significant optical losses. As a result, maintaining sufficient signal-to-noise ratio typically requires either higher incident optical power or longer camera exposure times. In practice, increasing optical power improves measurement fidelity but accelerates thermal decorrelation of the speckle pattern, while reducing exposure time or input power limits the available photon budget and degrades measurement accuracy. These competing effects constrain operation in photon-limited regimes and limit the accessible operating range of speckle-based sensing systems.

Reducing measurement noise in speckle metrology could help alleviate these constraints; however, this task is inherently challenging because the dominant noise sources are fundamentally intertwined with the interference process itself. In conventional imaging, speckle is typically regarded as noise that can be separated from the underlying structure of the scene. In contrast, in speckle metrology, the speckle pattern directly encodes the measurement information. Consequently, many existing denoising approaches are designed to completely suppress or eliminate speckle \cite{SpeckleDenoisingDHM,SpeckleDenoisingOCT,SpeckleDenoisingDermoscopy,SpeckleNoiseReductionUltraSound}, making them unsuitable for metrological applications where preservation of the interference structure is critical. Similarly, standard denoisers trained on natural images often fail to distinguish between stochastic measurement noise and physically meaningful speckle grains, leading to degradation of the encoded signal.

Recent advances in generative modelling—particularly denoising diffusion probabilistic models (DDPMs)—have demonstrated a strong ability to learn complex image distributions and achieve high-fidelity denoising \cite{DDPM_paper,SohlDickstein,Nichol2021}. In this work, we adapt the real-world image denoising framework of Yang et al. \cite{real_world_denoising} for application in speckle metrology. Further details on DDPMs, their implementation, and the network architecture are provided in Section \ref{methods_section}.
While diffusion models have been successfully applied to optical imaging tasks such as super-resolution microscopy and image reconstruction \cite{volpe_2024,li_microscopy_2024}, their use in precision optical metrology has not yet been reported, representing a key contribution of this work.
Despite their promise, diffusion-based approaches raise important challenges, particularly regarding generalisation and the potential introduction of hallucinated features \cite{li_microscopy_2024,he2025diffusion}. Additionally, assembling large, labelled experimental datasets is often impractical in speckle metrology, as thermal decorrelation can lead to label drift over time \cite{Facchin2024}.
To address these limitations, we introduce a transfer learning strategy tailored to optical speckle metrology. The model is first pre-trained on simulated data and subsequently fine-tuned on a small experimental dataset to capture system-specific correlations \cite{Kok:25}. This approach improves robustness while mitigating data scarcity and domain-shift effects.

%Recent advances in generative modelling—particularly denoising diffusion probabilistic models (DDPMs)—have demonstrated strong capability in learning complex image distributions and achieving high-fidelity denoising \cite{DDPM_paper,SohlDickstein,Nichol2021}. In this work, we adapt the real-world image denoising framework of Yang et al. \cite{real_world_denoising} to speckle metrology. Further details on DDPMs, their implementation, and the network architecture are provided in Section \ref{methods_section}. While diffusion models have been applied in optical imaging tasks such as super-resolution microscopy and image reconstruction \cite{volpe_2024,microscopy_denoising}, their use in precision optical metrology has not yet been reported and is thus a key innovation of this paper.

%There are potential issues to this approach, partly due to concerns about generalisation and hallucinated features \cite{microscopy_denoising,he2025diffusion}. In addition, large labelled experimental datasets are difficult to obtain, as thermal decorrelation can cause label drift over time \cite{Facchin2024}. To address these challenges for optical metrology, we employ transfer learning for the first time to this field: the model is pre-trained on simulated data and then fine-tuned on a small experimental dataset to capture system-specific correlations \cite{Kok:25}.

We introduce a diffusion-based denoising framework tailored to speckle metrology as an AI-enabled sensing paradigm. Unlike conventional denoising approaches, it effectively suppresses measurement noise while preserving the intrinsic speckle structure and encoded information. We validate the framework using speckle patterns captured from an integrating sphere under low-exposure conditions, with controlled variations in the incident wavelength. Principal component analysis (PCA) of the denoised data demonstrates a substantial reduction in measurement uncertainty compared to solely using the raw images as data. This approach enables reliable measurements under constrained illumination and requires only a single calibration step for new datasets. It is particularly suited to applications such as low-power laser stabilisation and operation in spectral regions with limited optical power or detector sensitivity, including the deep-ultraviolet and mid-infrared regimes\cite{peterkovic_optimizing_2025,pret_evidence_2012,hajjarian_tutorial_2020}.

\section{Results}
\begin{figure*}
    \centering
    \includegraphics[width=1\textwidth]{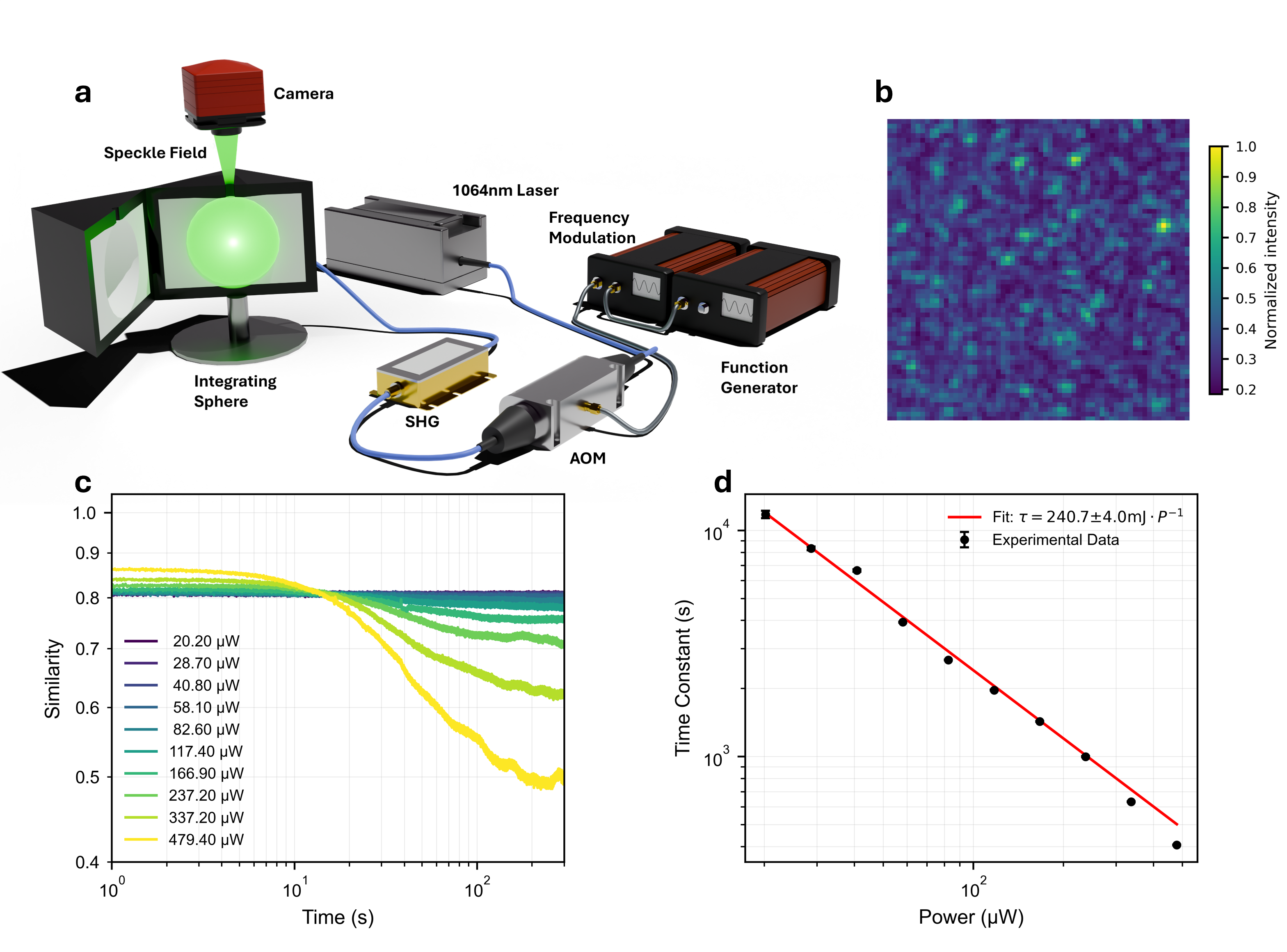}
    \caption{\textbf{a.} A simplified diagram of the experimental setup. A 1064nm laser is doubled to 532nm via second-harmonic generation (SHG) and wavelength modulated with an acousto-optic modulator (AOM) before being coupled into an integrating sphere. The resulting speckle pattern is captured by a camera for analysis. Further details of the setup may be found in Section 4.2. \textbf{b.} An example of a high-signal speckle frame at a resolution of 64x64 pixels. \textbf{c.} Temporal evolution of speckle similarity without applied modulation. At higher powers, the speckle pattern changes rapidly over time, leading to significant decorrelation on experimentally relevant timescales. Reducing optical power improves temporal stability but reduces short-term similarity and therefore measurement fidelity. \textbf{d.} The decorrelation times is inversely proportional to the optical input power with $\tau = 240.7\mathrm{mJ}/P$.}
    \label{fig:motivation}
\end{figure*}

Thermally induced decorrelation is a key limiting factor in integrating-sphere-based measurements, as illustrated in Fig. \ref{fig:motivation}. As the injected optical power increases, the temporal stability of the speckle pattern degrades, leading to a more rapid decay in correlations between the patterns. Figure \ref{fig:motivation}a shows the experimental arrangement used to generate speckle patterns using an integrating sphere. Light entering the sphere undergoes multiple scattering events, producing a complex interference pattern at the detector, as shown in Fig. \ref{fig:motivation}b. This speckle pattern is highly sensitive to changes in system parameters, enabling precision optical metrology. Figure \ref{fig:motivation}c shows the temporal evolution of speckle similarity for different incident power levels.
For our specific integrating sphere, we observe a clear inverse relationship between the decay time constant $\tau$ and the incident power $P$, described by $\tau = \qty{240.7}{mJ}/P$ in Fig. \ref{fig:motivation}d. This suggests that to improve the stability, one must reduce the optical power. Notably, the similarity metric reaches unity only in the absence of measurement noise, which inherently biases high-sensitivity measurements toward operation at elevated power levels.
This behaviour introduces a fundamental trade-off between measurement resolution and temporal stability: higher optical power improves signal fidelity but increases the rate of thermal decorrelation, thereby reducing the achievable measurement duration.
%Thermally induced decoherence is established as a limiting factor in integrating-sphere-based measurements, as illustrated in Fig. \ref{fig:motivation}. As the injected optical power into the sphere increases, the speckle pattern similarity decays more rapidly over time. Fig. \ref{fig:motivation}c. presents the temporal evolution of the speckle similarity for different incident power levels. For our specific integrating sphere, we observe a clear inverse relationship between the decay time constant $\tau$ and the incident power $P$, quantitatively described by $\tau = \qty{240.7}{mJ}/P$. It should be noted that the similarity metric can only reach unity in the absence of measurement noise, which biases more sensitive measurements toward operation at higher power levels. This, in turn, introduces a fundamental trade-off between measurement resolution and temporal stability.

%To enable operation in this low-power regime without degrading measurement performance, we first introduce a denoising step to recover the underlying speckle morphology from the unwanted noise.  The denoised speckle frames are then used in place of the raw images in the standard analysis pipeline, as illustrated in Fig. 2a. Details of the denoising procedure are described in Section 4.1.

\begin{figure*}
    \centering
    \includegraphics[width=1\textwidth]{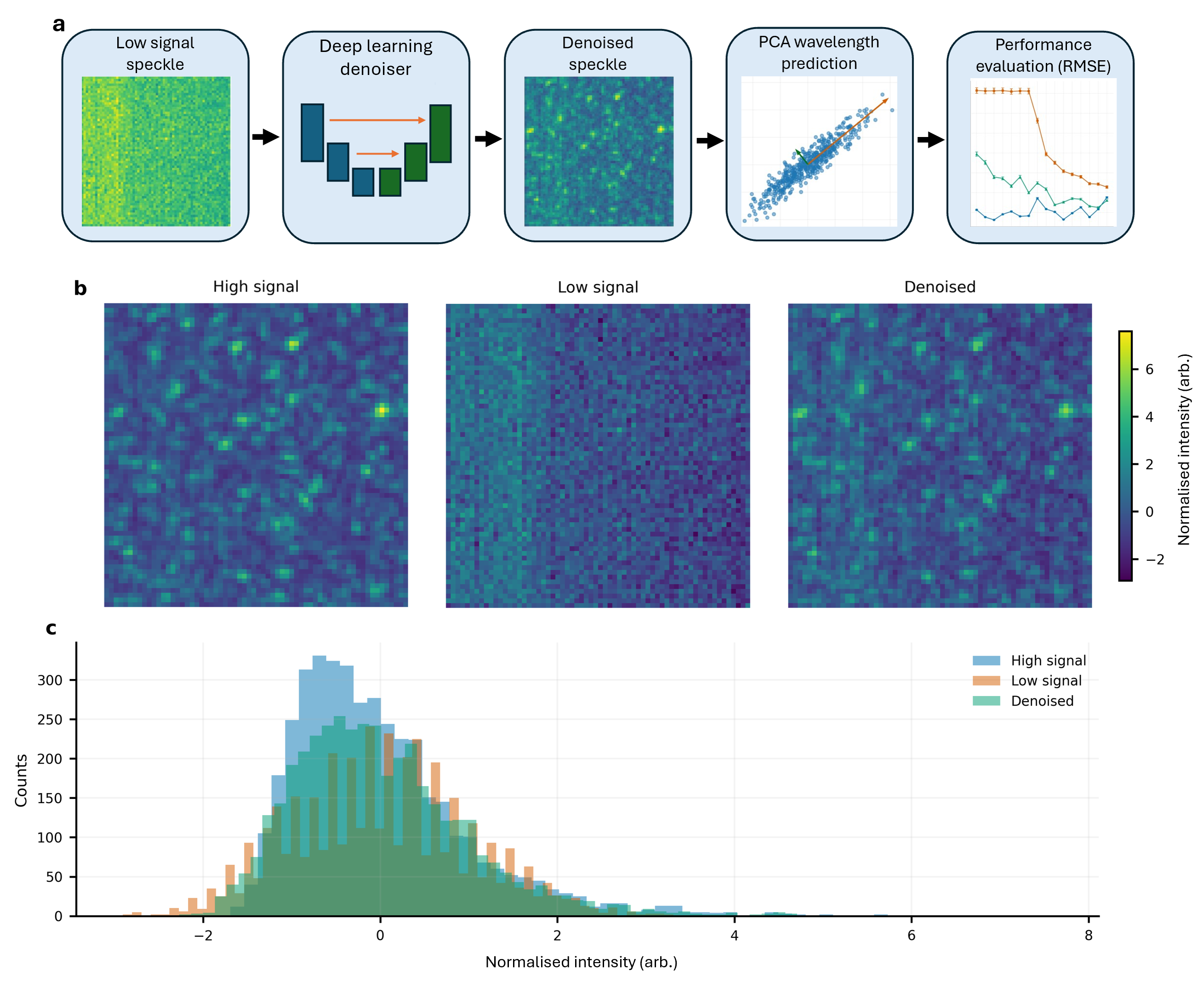}
    \caption{\textbf{a.} A high-level overview of the denoising and wavelength prediction process. The low-signal speckle is processed through the denoiser to produce a denoised speckle frame. PCA is then applied to the denoised frame for wavelength predictions. Lastly, the performance of this prediction is evaluated. \textbf{b.} Comparison of representative speckle frames from the high-signal, low-signal, and denoised at an exposure time of \qty{50}{\micro s} for the low-signal and denoised frames. The denoised frame exhibits speckle grain structures that more closely resembles those of the HS frame, in contrast to the LS frame where the grain structure is not observed. All images are normalised on a frame by frame basis, highlighting the underlying structure of each frame, rather than comparing their true pixel values. \textbf{c.} Normalised pixel-value histograms for the three frames shown in panel b. The denoised distribution more closely resembles the high-signal distribution, while the low-signal frame shows discrete quantisation peaks due to overwhelming detector noise in a photon-starved regime.}
    \label{fig:speckle_frame_comparisions}
\end{figure*}

To assess whether the denoising process uncovers the underlying speckle structure, representative speckle frames from the low-signal (LS), high-signal (HS), and denoised datasets are shown in Fig. \ref{fig:speckle_frame_comparisions}b. The denoised frame reveals grain morphology and spatial structure consistent with the HS frame, whereas in the LS frame the speckle contrast is reduced and the grain structure is largely obscured by noise, limiting its use for subsequent analysis. Consistent with this, the distribution of pixel intensities in the denoised frame also aligns with that of the HS data, in contrast to the LS case (Fig. \ref{fig:speckle_frame_comparisions}c). 
%The denoised frame exhibits grain structures that closely resemble those of the HS frame, in contrast to the LS frame where the structure is obscured by noise. 
%The absolute difference between the denoised and HS frames highlights residual discrepancies, primarily associated with reduced contrast in some grains. 

Although the reconstruction is not exact, the preservation of the dominant speckle features and spatial correlation structure indicates that information relevant to the standard linear retrieval pipeline is retained.%supports the observed reduction in wavelength estimation error obtained using an unchanged linear retrieval pipeline.

\begin{figure}[!htb]
    % Single column figure
    \centering
    \includegraphics[width=1\columnwidth]{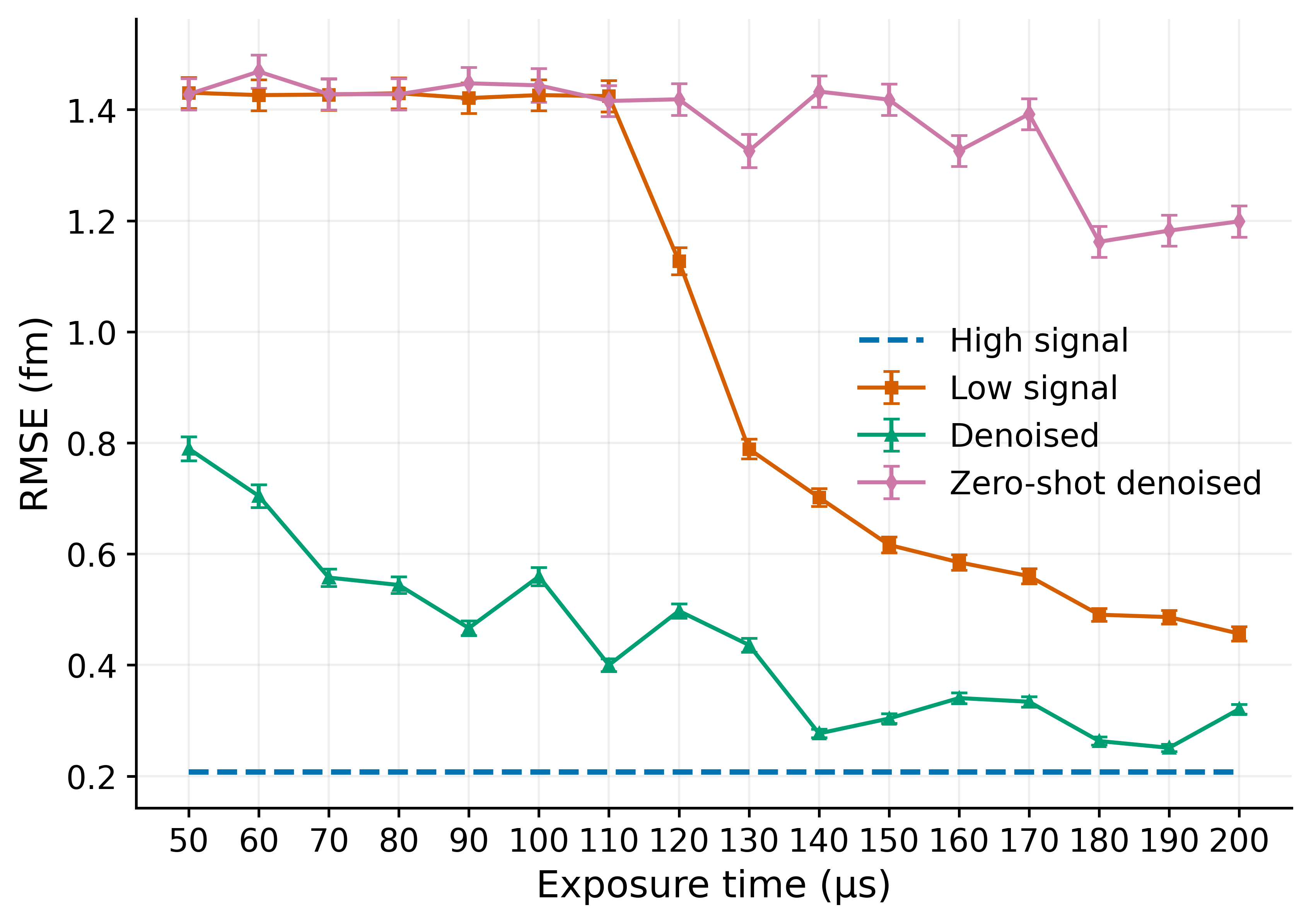}
    \caption{Root mean square error (RMSE) of the wavelength predictions as a function of exposure time for LS and denoised speckle datasets. Results are shown for the 5-epoch fine-tuned denoiser. The fine-tuned denoiser yields a substantial reduction in RMSE at low exposure times, with diminishing improvement as signal levels increase. RMSE values are averaged over five independent inference runs for each dataset. Error bars indicate the standard deviation across inference runs. The RMSE for HS speckle is also shown as a reference baseline, representing the performance obtained at high signal levels, and is averaged across all data runs due to the data collection method described in Section 4.2.}
    \label{fig:speckle_rmse}
\end{figure}

Having established that the denoised images preserve relevant speckle structure, we next assess their impact on wavelength estimation accuracy. Exposure time is varied to control the photon count in the recorded speckle patterns, with shorter exposures corresponding to the photon-starved regime identified in Fig. \ref{fig:motivation}. HS corresponds to a 3000 $\mu$s exposure and therefore is the high-signal reference condition. Fig. \ref{fig:speckle_rmse} summarises the denoising performance across all investigated exposure times by reporting the root mean square error (RMSE) of the wavelength predictions obtained from LS, denoised, and HS speckle datasets. Initial analysis using the baseline denoiser provided marginal improvement, indicating that pre-training on simulated speckle alone is insufficient to recover the measurement-relevant correlations present in the experimental data, necessitating fine-tuning on experimental data, which enables the improved performance observed in Fig. \ref{fig:speckle_rmse}.

\begin{figure*}[!htb]
    \centering
    \includegraphics[width=1\textwidth]{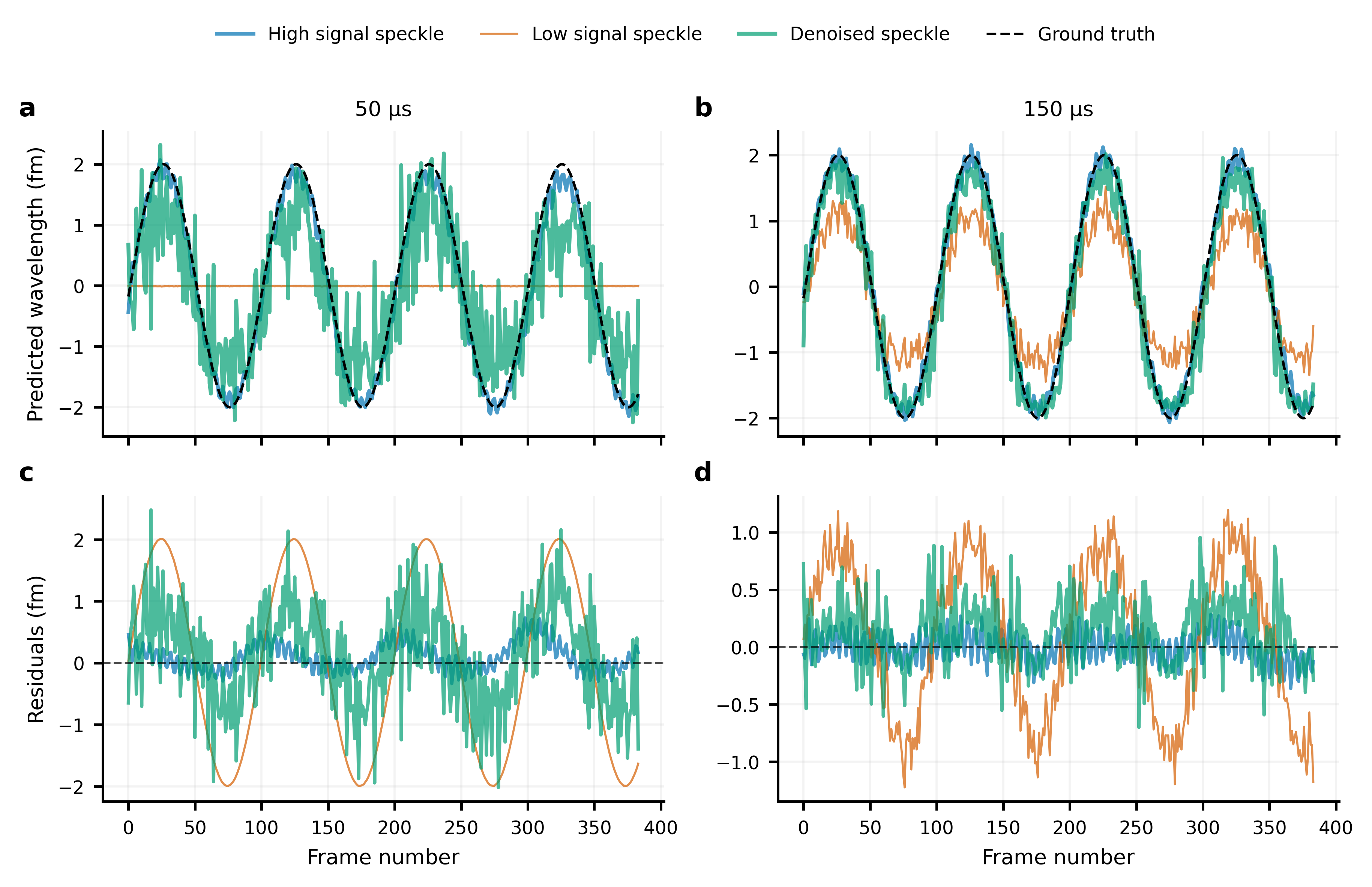}
    \caption{Wavelength predictions and the corresponding residuals derived from speckle datasets acquired at exposure times of \textbf{a.} \qty{50}{\micro s} and \textbf{b.} \qty{150}{\micro s}. For each exposure time, results are reported for both the LS speckle frames and the denoised speckle frames. The denoised speckle frames were generated using a denoising model that was fine-tuned for 5 training epochs. At \qty{50}{\micro s}, the LS data fails to reliably reproduce the imposed sinusoidal wavelength modulation, whereas application of the denoising procedure enables an accurate reconstruction of the modulation. At \qty{150}{\micro s}, denoising further decreases the prediction uncertainty and enhances the agreement with the imposed modulation. Panels \textbf{c.} and \textbf{d.} display the prediction residuals associated with \textbf{a.} and \textbf{b.}, respectively.}
    \label{fig:50usV100usWavelength}
\end{figure*}

For the lowest exposure times (\qtyrange{50}{110}{\micro s}), the fine-tuned denoiser yields a substantial reduction in RMSE relative to the LS datasets, with a reduction of 44.8\%, 50.61\%, 60.94\%, 61.93\%, 67.16\%, 60.77\% and 71.92\% respectively. Over this exposure-time range, Fig. \ref{fig:speckle_rmse} shows that the RMSE of the LS speckle data remains approximately constant at 1.43 fm. This is because the wavelength prediction algorithm fails to reconstruct the applied sinusoidal modulation of the incident light and instead predicts a constant wavelength. An example of this behaviour is provided in Fig. \ref{fig:50usV100usWavelength}a, where the LS wavelength reconstruction exhibits no discernible modulation. As the LS prediction is approximately constant, the residuals are dominated by the imposed sinusoidal wavelength modulation. For a sinusoid, the RMSE is related to the amplitude by a factor of $\frac{1}{\sqrt{2}}$, hence for an amplitude of 2 fm the expected RMSE is $2/\sqrt{2} \approx 1.41fm$, consistent with the observed RMSE in Fig. \ref{fig:speckle_rmse}. Across intermediate exposure times, the fine-tuned model continues to outperform the LS datasets, although the magnitude of improvement decreases as signal increases. At the highest exposure times, the RMSE values of the LS and denoised datasets begin to converge, consistent with the increased signal-to-noise ratio reducing the benefit of denoising. However, the denoised datasets continue to outperform the LS datasets. Overall, our results indicate that fine-tuning is necessary to decrease the RMSE of our wavelength prediction based on speckle analysis. Furthermore, only a small number of additional iterations are required for fine-tuning to achieve a substantial improvement, as can be seen in Fig. \ref{fig:speckle_rmse}. 

Fig. \ref{fig:50usV100usWavelength} presents representative wavelength predictions and residuals for the exposure time data sets \qty{50}{\micro s} and \qty{150}{\micro s}. At \qty{50}{\micro s}, the PCA of the LS dataset fails to recover the applied sinusoidal wavelength modulation and predicts a constant wavelength, whereas denoising enables clear reconstruction of the modulation with substantially reduced residuals. At \qty{150}{\micro s}, partial recovery of the modulation is already visible in the LS dataset; however, denoising further reduces the uncertainty and yields predictions that more closely match the applied modulation. For $50 \mu s$, the RMSE is reduced from 1.43 fm to 0.79 fm and for 150 $\mu s$ the RMSE is reduced from 0.62 fm to 0.31 fm.

\section{Discussion}
The results demonstrate that diffusion-based denoising can effectively suppress measurement noise in speckle frames while preserving the physically meaningful structure required for precision metrology. The reduction in wavelength prediction error at low exposure times indicates that the denoiser is not merely producing visually plausible speckle patterns, but is retaining the correlations necessary for accurate linear analysis, which underpins the observed improvement in measurement accuracy. The diminishing improvement at higher exposure times is consistent with increasing signal-to-noise ratio reducing the need for algorithmic assistance. Our method therefore appears particularly well suited to low-illumination, photon-starved acquisition regimes.

A key finding is that effective denoising was achieved only after fine-tuning the model on experimentally labelled data. While pre-training on simulated speckle provides a useful prior on generic speckle statistics, it does not capture the system-specific correlations imposed by the experimental configuration. Fine-tuning enables the model to adapt to these correlations using a comparatively small dataset, allowing us to use only 800 speckle frames for further training, reducing overhead as the model is pretrained once on simulated data for 200 epochs, with subsequent calibrations and fine-tuning requiring only 5 additional epochs. This highlights the importance of combining physics-informed simulation with experimental data when applying generative models to metrological tasks \cite{karniadakis_physics-informed_2021}.

%\textbf{This reducing both training time and calibration overhead by approximately a factor of 20. Instead of training directly on experimental data for 200 epochs during each calibration, the model is pretrained once on simulated data for 200 epochs, with subsequent calibrations and fine-tuning requiring only 5 additional epochs. }

The failure of a zero-shot denoiser to improve measurement performance further emphasises this point. Without fine-tuning, the diffusion model behaves primarily as a speckle image generator, producing visually plausible but uncorrelated frames \cite{li_microscopy_2024, linguraru_hallucination_2024}. In a metrological context, such behaviour is insufficient, because the encoded measurement information resides in subtle correlations between speckle patterns \cite{Facchin2024}. These results underscore the necessity of experimental calibration when applying generative denoising methods to information-bearing interference phenomena.

%The use of a strictly linear analysis pipeline strengthens confidence that the denoising procedure preserves the underlying measurement information. Specifically, the use of principal component analysis and linear regression ensure that no nonlinear feature extraction is introduced at the analysis stage. Improved performance following denoising therefore suggests that relevant physical information is retained rather than artificially introduced. Nonetheless, generative models inherently carry a risk of hallucinating features not present in the original data \cite{linguraru_hallucination_2024,li_microscopy_2024}. Visual inspection of denoised frames, together with consistency checks against independent physical measurements, therefore remains important.

A key question is whether the denoising process recovers the underlying speckle structure or introduces artificial correlations. The use of a strictly linear analysis pipeline strengthens confidence that the denoising procedure preserves the underlying measurement information. As the principal component analysis and linear regression steps operate within a strictly linear framework, the analysis stage cannot introduce nonlinear feature extraction. Any improvement in prediction accuracy must therefore arise from changes in the correlations present in the denoised data. The inability of the low-signal measurements to recover the modulation, combined with the successful reconstruction following denoising, indicates that the relevant correlations are restored rather than artificially introduced. While generative models inherently carry a risk of hallucinating features not present in the original data \cite{linguraru_hallucination_2024,li_microscopy_2024}, any such artefacts must still preserve the linear relationships exploited by the downstream analysis to yield improved performance. This places strong constraints on admissible reconstructions. The requirement for fine-tuning on experimental data further suggests that the model captures system-specific correlations rather than generating generic speckle-like structure. This interpretation is further supported by consistency with the applied wavelength modulation and agreement with high-signal measurements, providing independent validation that the recovered structure reflects the underlying physical behaviour.

Several practical considerations could further improve the present approach. Diffusion models incur non-negligible inference times, which scale with image size and the number of diffusion steps. More compact models could be deployed on field-programmable gate arrays (FPGAs) or other edge hardware for near real-time inference \cite{zhang2015optimizing, wang2016dlau, shawahna2018fpga}. In this work, inference was feasible because the images were limited to $64\times64$ pixels with 12-bit grayscale encoding, both of which reduce memory requirements per speckle frame. Extending the approach to larger images or more complex acquisition schemes would require careful consideration of computational constraints. In addition, the denoiser provides less benefit as signal levels increase, indicating that its primary utility lies in photon-limited sensing regimes.

 A current limitation of this work is the requirement that the 
underlying grain structure in both the low and high signal speckle remains sufficiently correlated for the training and test datasets. Thus, future work may focus on building a general speckle denoiser that would not require experimental data for a fine-tuning step. Such an approach would likely reduce the minimum signal level at which speckle can be reliably captured while still enabling accurate wavelength reconstruction.

A key implication of this work is that accurate speckle-based measurements can be performed at reduced signal levels, alleviating the trade-off between the signal-to-noise ratio and speckle stability. While increased optical power improves measurement fidelity, it also accelerates thermal decorrelation of the speckle pattern, limiting the usable measurement window. The denoising framework demonstrated here enables useful measurements at lower energy per frame by recovering the correlations required for accurate analysis in photon-limited conditions. As a result, the usable operating regime of speckle metrology is extended to shorter exposure times, lower input intensity, or increased sampling rates. Such capability is especially relevant for low-power lasers, sensing in spectral regimes where detector quantum efficiency is poor, and applications involving lossy delivery paths or photosensitive samples \cite{peterkovic_optimizing_2025, pret_evidence_2012, hajjarian_tutorial_2020}. In these settings, measurement performance is often constrained by photon budget, making data-driven, energy-efficient sensing strategies particularly valuable.

% \section{Conclusion}
We have demonstrated that a diffusion-based denoising framework, trained on simulated speckle data and fine-tuned using a limited set of experimentally labelled measurements, can recover the correlations required for accurate wavelength prediction from low-signal speckle measurements. The approach operates without modification to the downstream analysis, and its effectiveness within a strictly linear pipeline provides strong evidence that the underlying measurement information is preserved rather than artificially introduced. Substantial reductions in prediction error are achieved across the photon-limited regime, while requiring only a small amount of additional experimental data for calibration.

%can effectively recover metrological information from LS speckle frames. Applied to wavelength sensing using an integrating sphere, the method achieves an average of 59.73\% reduction over the \qtyrange{50}{110}{\micro s} exposure range when compared to unprocessed LS data, while preserving the relevant speckle correlations required for linear analysis. The approach requires only limited fine-tuning when adapting to a new dataset, substantially reducing experimental overhead.

These results show that tailored generative denoising can extend the usable operating regime of speckle-based metrology under constrained illumination conditions, without modifying the downstream analysis pipeline. As such, diffusion-based denoising provides a practical route to improving measurement fidelity in photon-limited speckle experiments.
 
\section{Methods}\label{methods_section}
\subsection{Denoising algorithm and training procedure}
We employ a DDPM to recover HS speckle structure from LS measurements. In conventional DDPMs, the forward process progressively perturbs a clean image with Gaussian noise over many iterations according to a variance schedule $\beta_t$ \cite{DDPM_paper}, until the image consists only of pure Gaussian noise. In this work, however, we do not diffuse toward pure Gaussian noise. Instead, following Yang et al. (2024) \cite{real_world_denoising}, we train the model using experimentally relevant pairs of HS and LS speckle frames as the initial and end points of the forward process. This ensures that the denoising trajectory remains confined to the physical noise regime of the measurement.

This modified forward diffusion process interpolates between a HS speckle frame $x_0$ and its corresponding LS measurement $x_T$ according to

\begin{equation}\label{labeled_fp}
    x_t = (1 - \alpha_t)x_0 + \alpha_t x_T,
\end{equation}

where $\alpha_t=t/T$, $t$ is the diffusion timestep, and $T$ is the total number of steps. In this formulation, $x_T$ represents the experimentally captured noisy frame, and the diffusion process spans only the range of degradations relevant to the measurement.

The reverse process estimates the denoised image by iteratively removing the noise component associated with each timestep:

\begin{equation}
    x_{t-1} = x_t - \frac{1}{T}[x_T-S_\theta(x_t,t)],
\end{equation}

where $S_\theta$ is a neural network conditioned on both the image $x_t$ and the timestep $t$. Conditioning on $t$ ensures that the magnitude of noise removal at each step mirrors the degradation applied during training, improving stability and convergence for highly noise-corrupted speckle frames. The neural network, $S_\theta$, is implemented as a 2D UNet model using the Hugging Face Diffusers library \cite{von-platen-etal-2022-diffusers}. The architecture is based on the original UNet proposed by Ronneberger et al. \cite{ronneberger}, with the inclusion of additional attention layers to allow the model to capture long range spatial dependencies within the speckle frames. A high level overview of this model architecture is depicted in Fig. \ref{fig:experiment}a.

To initialise the model, we generate 100 000 simulated speckle frames using a procedure previously outlined by Goodman \cite{GoodmanSimulatingSpeckle} with grain sizes chosen to closely match those observed experimentally. The simulated frames resembled our HS frames, and were degraded to form LS counterparts by applying additive Gaussian noise and scaling the intensity to 30\% of the original signal. This pre-training stage provides a physics-informed prior on speckle statistics without requiring a large experimentally labelled dataset.

The model was pre-trained for 200 epochs with a batch size of 256, a learning rate of $10^{-3}$, and $T=50$ diffusion steps. The loss function combined the structural similarity index measure (SSIM) and mean squared error (MSE),

\begin{equation}
    \mathcal{L} = [1-\mathrm{SSIM}(x,f_\theta(x))] + a\mathrm{MSE}(x,f_\theta(x)),
\end{equation}

where $a=10^{-3}$ for the first 10 epochs and is subsequently set to zero. This staged loss prevents early training instability in high noise regimes while favouring structure preserving convergence at later stages. It is important to note, that when fine-tuning this model, the MSE term remains included in the loss function, as fine-tuning required less than 10 epochs. 

Following pre-training, the model was fine tuned using experimentally labelled HS-LS speckle pairs acquired at different exposure times while the wavelength was sinusoidally modulated. Fine-tuning was carried out for 5 epochs with dropout enabled to reduce overfitting \cite{JMLR:v15:srivastava14a}, which is more likely given the limited size of the available training data. This transfer learning strategy enables adaptation to system specific speckle correlations using 800 experimental frames, substantially reducing the amount of labelled experimental data required relative to training from scratch. In addition, 200 frames were used for validation and 480 frames for testing. Since the exact speckle intensity pattern is specific to the optical configuration and can drift over time, rapid calibration is important for practical deployment in sensing applications \cite{basak_review_2012}.

\subsection{Experimental setup}
\begin{figure*}[!htb]
    \centering
    \includegraphics[width=\textwidth]{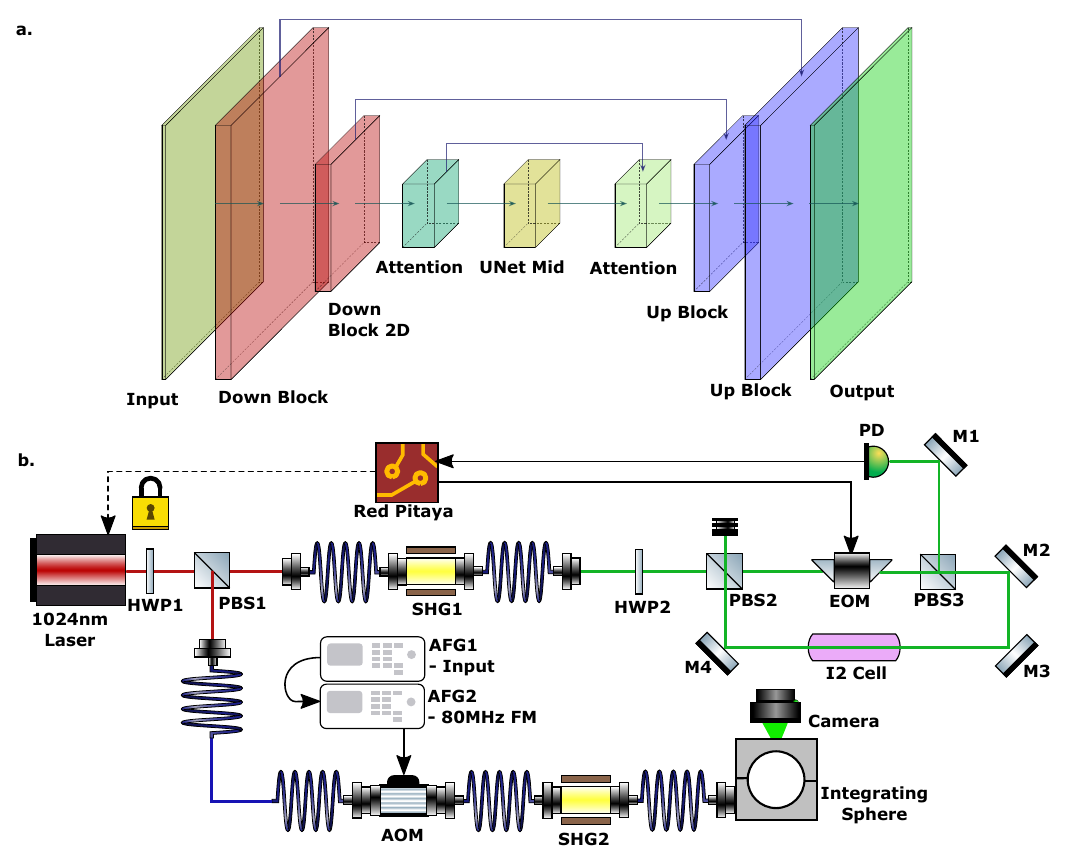}
    \caption{\textbf{a.} High-level overview of the diffusion model architecture used for speckle denoising. This model was implemented using the Hugging Face Diffusers library UNet2DModel class. Hence, each constituent block type abstracts away a combination of layers and operations, rather than a single neural-network layer. \textbf{b.} Schematic of the experimental setup. A frequency-stabilised 1064 nm laser is wavelength modulated using an acousto-optic modulator (AOM) and frequency doubled to produce light at 532nm via second harmonic generation (SHG). The modulated 532nm beam is coupled into an integrating sphere to generate a highly sensitive speckle pattern, which is imaged onto a high-speed monochrome camera for subsequent analysis. The frequency stabilisation is the result of an iodine frequency lock employing modulation transfer spectroscopy to lock to a hyperfine transition.}
    \label{fig:experiment}
\end{figure*}

Speckle patterns were generated using a 50 mm diameter integrating sphere with a high reflectance diffuse coating (Thorlabs 2P3/M), chosen to enhance speckle sensitivity through multiple scattering. The integrating sphere has an exit port diameter of \qty{11.5}{mm} and internal reflectance of 97\% at 532 nm. A frequency-stabilised  laser (Mephisto, Coherent) operating at a wavelength of 1064 nm was frequency modulated and subsequently frequency-doubled to produce 4 mW of 532 nm wavelength radiation. The use of a shorter wavelength improves speckle sensitivity by allowing a greater relative phase shift to accrue between paths \cite{Facchin2024} and enables detection using a visible range high-speed (\qty{935}{fps} at maximum resolution, up to \qty{80000}{fps} for reduced frame size), monochrome sCMOS camera (MV4-D1280-L01-GT, PhotonFocus). A detailed schematic of the experimental setup is shown in Fig. \ref{fig:experiment}b.

Wavelength modulation was applied using an acousto-optic modulator (AOM) driven by a function generator (Keysight) producing a frequency-modulated 80 MHz radio-frequency signal. A second function generator was used to drive the sinusoidal frequency modulation resulting in a peak-to-peak wavelength deviation of 4 fm at a modulation frequency of 3 Hz. The modulated light was coupled into the integrating sphere via a single mode, polarization maintaining fibre. This is critical for producing a stable speckle pattern as the resulting pattern is sensitive to both input polarisation and wavefront structural modes \cite{aman2026speckle, facchin_speckle_based_2020, ambichl_super_2017, carpenter_observation_2015}. It should be noted that wavelength modulation via the AOM introduces a power modulation. The speckle frames are individually normalised upon collection ensuring zero mean and unit variance, thereby removing the overall frame-by-frame brightness variation.

To generate labelled datasets for training and evaluation, speckle frames were acquired sequentially at alternating exposure times. Datasets were collected in pairs: first, a high signal reference was recorded at an exposure time of \qty{3000}{\micro s} followed immediately by a LS dataset, together forming a single dataset pair. The LS dataset exposure times ranged from \qtyrange{50}{200}{\micro s} in increments of \qty{10}{\micro s}. This alternating acquisition strategy ensures that the underlying speckle grain structure remains consistent between high and low signal measurements while mitigating thermal effects within the integrating sphere causing the speckle grain structure to slowly drift. 

\subsection{Speckle analysis}
To extract wavelength information from the speckle measurements, all datasets were analysed using an identical, linear processing pipeline based on PCA followed by linear regression. PCA is a widely adopted technique in speckle-based metrology for dimensionality reduction and noise suppression, and preserves the dominant linear correlations within the data \cite{pearson_PCA,Metzger2017,aman2026speckle,bruce_overcoming_2019}. The application of PCA allows for an unsupervised method of extracting the linear projection of the measurement. With a dimensionally reduced dataset, the required degrees of freedom in the supervised linear regression model can be significantly reduced, depending on speckle image size.

PCA was fitted separately to the LS, HS and denoised datasets, with the first three principal components being retained. These components were then used as inputs to a linear regressor to predict the applied wavelength modulation. Both PCA and regression were implemented using the scikit-learn library \cite{scikit-learn}. The dimensionality reduction from the PCA pre-processing step was performed to filter some random measurement noise and reduce the numbers of degrees of freedom for the linear regressor to fit. The linear regressor provides a supervised and linear means of wavelength prediction.

Crucially, the analysis pipeline was not modified between the HS, LS, and denoised datasets. As PCA and linear regression operate entirely within a linear framework, successful recovery of the wavelength modulation from denoised speckle frames is a strong indicator that the denoising process preserves the physically meaningful structure of the speckle pattern rather than introducing artificial correlations.

To quantify performance, the RMSE between the predicted and applied wavelength modulation was calculated for each dataset. RMSE values were averaged over 5 independent inference runs per exposure time to mitigate stochastic variability in the denoising process. These results are presented in Fig. \ref{fig:speckle_rmse}. No nonlinear feature extraction or model-based fitting was applied at any stage of the analysis. This constraint was imposed to limit the number of degrees of freedom within the model, and consequently reduce the risk of over-fitting. %\textcolor{red}{why si this important and how might it lead to erroneous results if we did?....}
% \textcolor{red}{how was the model tested to ensure overfitting did not occur?} \textbf{I WILL INCLUDE SOME VALIDATION VS LOSS CURVES. -ADMIR}

\begin{backmatter}
\bmhead{Acknowledgements}
We thank the Australian Research Council for funding (grant FL210100099). This research was partially funded by the Australian Research Council Industrial Transformation Training Centre for Current and Emergent Quantum Technologies (IC240100012). This research was supported by an Australian Government Research Training Program (RTP) Scholarship. This work is supported by the U.S. Office of Naval Research Global (N62909-25-1-2029). %A. Bajraktarevic 
We would like to thank Gabriel Britto Monteiro and Chris Perrella for experimental assistance and critical comments on the manuscript. We acknowledge the contribution of Lewis McMillan to early stages of this project.

\bmhead{Disclosures}
The authors declare no conflicts of interest.

\bmhead{Data availability}
Data used to generate the figures are deposited onto Figshare, and can be accessed with the following URL:  https://figshare.com/s/2c1d19b372d3e11c03f5.
\end{backmatter}

%%%%%%%%%%%%%%%%%%%%%%% References %%%%%%%%%%%%%%%%%%%%%%%%%
\bibliography{refs}

\end{document}